\documentclass[epj]{webofc}
\usepackage[utf8]{inputenc}
\usepackage[varg]{txfonts}   
\usepackage{booktabs}
\usepackage{xcolor}
\definecolor{darkred}{rgb}{0.4,0.0,0.0}
\definecolor{darkgreen}{rgb}{0.0,0.4,0.0}
\definecolor{darkblue}{rgb}{0.0,0.0,0.4}
\usepackage[bookmarks,linktocpage,colorlinks,
    linkcolor = darkred,
    urlcolor  = darkblue,
    citecolor = darkgreen]{hyperref}
\usepackage{subfigure}
\wocname{EPJ Web of Conferences}
\woctitle{Lattice2017}
\DeclareMathOperator{\quarkl}{\mathcal{Q}}

\DeclareMathOperator{\smear}{\mathcal{S}}
\DeclareMathOperator{\prop}{K^{-1}}

\begin{document}
%
\selectlanguage{english}
\title{%
Towards extracting the timelike pion form factor on CLS two-flavour ensembles 
}
\author{%
\firstname{Felix} \lastname{Erben}\inst{1,2}\fnsep\thanks{Speaker, \email{erben@kph.uni-mainz.de}} \and
\firstname{Jeremy} \lastname{Green}\inst{3} \and
\firstname{Daniel}  \lastname{Mohler}\inst{1,2} \and
\firstname{Hartmut}  \lastname{Wittig}\inst{1}\fnsep
}
\institute{%
Institut für Kernphysik, Johannes Gutenberg-Universität Mainz,
D-55099 Mainz, Germany
\and
Helmholtz Institut Mainz,
D-55099 Mainz, Germany
\and
NIC, DESY, Zeuthen, Germany
}
\abstract{%

Results are presented from an ongoing study of the $\rho$ resonance. The focus is on CLS 2-flavour ensembles generated using $\mathcal{O}(a)$ improved Wilson fermions with pion masses ranging from $265$ to $437$ $\mathrm{MeV}$. The energy levels are extracted by solving the GEVP of correlator matrices, created with the distillation approach involving $\rho$ and $\pi\pi$ interpolators. The study is done in the centre-of-mass frame and several moving frames.   One aim of this work is to extract the timelike pion form factor after applying the L\"uscher formalism. We therefore plan to integrate this study with the existing Mainz programme for the calculation of the hadronic vacuum polarization contribution to the muon $g-2$.
}
\maketitle
\section{Introduction}\label{intro}

The $\rho$ resonance, whose principal decay is $\rho \rightarrow \pi\pi$, is the simplest QCD resonance to study on the lattice, yet very interesting for a number of reasons: Studying resonances is quite challenging and the $\rho$ is the benchmark for L\"uscher-style analyses \cite{luscher1,luscher2,luscher_mf}. One of the reasons for this is that all decays other than the already mentioned $\rho \rightarrow \pi\pi$ are negligible \cite{PDG-2016}. Also the noise-to-signal ratio, which is proportional to $e^{-(m_\rho-m_\pi)\Delta t}$ (where $m_\pi$ denotes the mass of a pion at rest), is more favourable compared to other hadronic states. Another reason to study the $\rho$ resonance is that it gives access to an interesting physical quantity: Meyer \cite{Meyer:timelike} has shown that one can extract the pion form factor $F_\pi$ in the timelike region for $2m_\pi \leq \sqrt{s} \leq 4m_\pi$, by computing scattering phase shifts and matrix elements in the vector channel. This quantity is of particular interest because it is crucial to reduce the uncertainty in theoretical calculations of hadronic vacuum polarization part $a_\mu^\mathrm{hvp}$ in the anomalous magnetic moment of the muon $(g-2)_\mu$ \cite{a_hvp}. One of the technical challenges of including multi-hadron interpolators is that we have to handle sink-to-sink quark propagators. The approach we use in this work and which is able to handle those is (stochastic) distillation using Laplacian-Heavyside (LapH) smearing \cite{dist1,dist2}.

\newpage
\section{Theoretical approach and lattice setup}\label{sec-1}

While there are efficient techniques for the inversion of the Dirac matrix $K$, it is still prohibitively expensive to compute the full propagator matrix. For a naive computation of a sink-to-sink quark line, this inversion would have to be performed for each pair of quark positions on the sink timeslice. Distillation drastically reduces the number of inversions needed by using a low-rank hermitian smearing matrix $\smear = V_S V_S^\dagger$ \cite{dist1}. This leads to a much smaller matrix, $V_S^\dagger K^{-1} V_S$, which has to be calculated and stored on disk. In stochastic LapH, which replaces the exact determination of the quark propagation by a stochastic estimate, quark lines $\quarkl$ (more precisely: quark propagators smeared at the source and sink) are then expressed as an expectation value of distillation-sink vectors $\varphi$ and distillation-source vectors $\varrho$:
\begin{align}
\nonumber \quarkl  &= \sum_b \langle \varphi^{[b]}(\rho) ( \varrho^{[b]}(\rho))^\dagger \rangle
,\\
\varphi^{[b]}(\rho) = &\smear \prop V_S P^{(b)} \rho , 
\phantom{aaaa}
\varrho^{[b]}(\rho)  = V_S   P^{(b)} \rho
.
\end{align}
The noise vectors $\rho$ are defined in the distillation subspace, obeying $\langle\rho\rangle=0$ and $\langle\rho \rho ^\dagger\rangle=1$. $P^{(b)}$ are the dilution projectors \cite{dilution} with dilution index $(b)$ in the distillation subspace.

In this setup we use $\rho$ interpolators and $\pi\pi$ interpolators with respective pion momenta $\mathbf{p}_1,\mathbf{p}_2$ obeying $\mathbf{p}_1+\mathbf{p}_2 = \mathbf{P}$,
\begin{align*}
\rho^0(\mathbf{P},t) =
\frac{1}{2 L^{3/2}} \sum_\mathbf{x} e^{-i \mathbf{P} \cdot \mathbf{x}} 
\bigg( \bar{u} \Gamma u - \bar{d} \Gamma d \bigg) (t) \phantom{a} , \phantom{a} \Gamma \in\ \lbrace \gamma_i, \gamma_0 \gamma_i \rbrace \phantom{a} ,
\end{align*}
\begin{align*}
(\pi \pi)(\mathbf{p}_1,\mathbf{p}_2,t) =
\pi^+(\mathbf{p}_1,t)
\pi^-(\mathbf{p}_2,t)
-\pi^-(\mathbf{p}_1,t)
\pi^+(\mathbf{p}_2,t)
,
\end{align*}
\begin{align}
\pi^+ (\mathbf{q},t) =
\frac{1}{2 L^{3/2}} \sum_\mathbf{x} e^{-i \mathbf{q} \cdot \mathbf{x}}
\big( \bar{u} \gamma_5 d \big) (\mathbf{x},t),
\phantom{aaa}
\pi^- (\mathbf{q},t) =
\frac{1}{2L^{3/2}} \sum_\mathbf{x} e^{-i \mathbf{q} \cdot \mathbf{x}}
\big( \bar{d}  \gamma_5 u \big) (\mathbf{x},t) \phantom{a} .
\end{align}
We work in the isospin limit with degenerate light quark masses $m_u = m_d$. Using these interpolators we analyse four different frames: The centre-of-mass frame (CMF) with a total momentum $\mathbf{P} = \frac{2 \pi}{L} \mathbf{d}=0$, as well as three moving frames with lattice frame momenta $\mathbf{d}^2 = {1,2,3}$, averaged over all possible directions on the lattice. In those frames we analyse different lattice irreducible representations (irreps).

In order to extract not only the ground state but also excited states, we use the variational method   \cite{gevp1,gevp2}: In this method, in each frame and irrep, we form a correlator matrix $C(t)$ from the interpolators defined above,
\begin{align*}
C(t) = 
\begin{pmatrix}
\langle  \rho(t) \rho^\dagger (0) \rangle  & 
\langle \rho(t) (\pi \pi)^\dagger (0) \rangle \\
\langle  (\pi \pi)(t) \rho^\dagger (0) \rangle &
\langle  (\pi \pi)(t) (\pi \pi)^\dagger (0) \rangle\\
\end{pmatrix} \phantom{a} .
\end{align*}
We then solve a generalised eigenvalue problem (GEVP) of this matrix,
\begin{align}
C(t) \mathbf{v} = \lambda(t) C(t_0) \mathbf{v} \phantom{a} .
\end{align}
There are different ways of choosing $t_0$. We use the so-called window method \cite{gevp_window}: $t_0 = t - t_w$, choosing a fixed window width of $t_w=3$. In order to assess residual excited-state-effects, we also employ the method where $t_0$ is kept at a fixed value for comparison. Asymptotically the eigenvalues satisfy $\lambda^{(k)}(t) \rightarrow e^{- E_k \, t_w}$, and we define effective masses in the usual way.

Using the L\"uscher condition \cite{luscher1,luscher2,luscher_mf}
\begin{align}
\delta(k) + \phi(q) = n \pi  \phantom{a} , \phantom{a}  k=\frac{2\pi}{L} q \phantom{a} ,
\end{align}
we can obtain information on the infinite-volume phase shift $\delta(k)$ from the discrete energy levels $E_{cm}=E_{cm}(k)=2 \sqrt{k^2+m_\pi^2}$ (boosted to the centre-of-mass frame) extracted from the finite-box lattice. 

In addition to the phase shift we are interested in the overlap of the vector current between the vacuum and the extracted states of the energy spectrum, $|A|=|\langle \Omega | J(t) | n \rangle|$ \cite{Meyer:timelike}, because it gives us access to the timelike pion form factor:
\begin{align}
|F_\pi(E)|^2 = G(\gamma) \left( q \frac{d \phi (q^2)}{dq} + k \frac{\partial \delta_1(k)}{\partial k}  \right) \frac{3 \pi E^2}{2 k^5} |A|^2  \phantom{a} ,
\end{align}
where $G(\gamma)$ is a factor of $\gamma$ or $\gamma^{-1}$, depending on the lattice irrep introduced in \cite{Max} and applied for the first time in \cite{aoki_2014}. In order to obtain this, we have calculated the matrix elements $\langle  J_\mu(t) O_i^\dagger(0) \rangle$.  Then, the GEVP eigenvectors $v_n(t)$ can be used to form operators $X_n(t) = v_n^\dagger(t) O(t)$ which project approximately onto the state with energy $E_n$. We can finally use these operators $X_n$ to form a two-point function with the current insertions at the sink,
\begin{align}
\langle J(t) X_n^\dagger(0) \rangle \rightarrow \langle \Omega | J(t) | n \rangle Z_n^* e^{-E_n t} \phantom{a} .
\end{align}
Forming various ratios \cite{Ben} the desired matrix element $|A|$ can then be extracted.

For this study, we use a Clover action on three different CLS 2-flavor lattices with $\beta=5.3$ and $c_{sw}=1.90952$ which corresponds to a lattice spacing of $a=0.0658(7)(7)\,\mathrm{fm}$ \cite{params_E5} and pion masses ranging from $437$ MeV to $265$ MeV \cite{params_E5_2}, listed in Table \ref{tab-ensembles}.
\newline

\begin{table}[thb]
  \small
  \centering
  \caption{CLS 2-flavor lattices used in this study. All lattices have $\beta=5.3$ and a lattice spacing of $a=0.0658(7)(7)\mathrm{fm}$. The numbers in brackets for $N_\mathrm{meas}$ are the target statistics in the respective ensembles.}
  \label{tab-ensembles}
\begin{tabular}{cccccccc}
\hline 
\hline 
 & T/a & L/a & $m_\pi$ [MeV]  & $\kappa$ & $m_\pi L$ & $N_\mathrm{conf}$ & $N_\mathrm{meas}$ \\ 
\hline 
E5 &64 & 32 &  437 &  0.13625 & 4.7 & 500 & 2000 \\ 
\hline 
F6 & 96 & 48 & 311 &  0.13635 & 5.0 & 300 & 300 (900) \\ 
\hline 
F7 & 96 & 48 & 265 &  0.13638 & 4.2 & 350 & 350 (1050) \\ 
\hline 
\hline 
\end{tabular} 
\end{table}

We use exact distillation (full dilution) on E5 for the quark lines connected to the source time\-slice and stochastic distillation (using time dilution) for the sink-to-sink lines. On F6 and F7 we use stochastic distillation for all lines.

\section{Analysis}\label{sec-2}

\begin{figure}[thb] 
  \centering
  \includegraphics[width=.9\linewidth,clip]{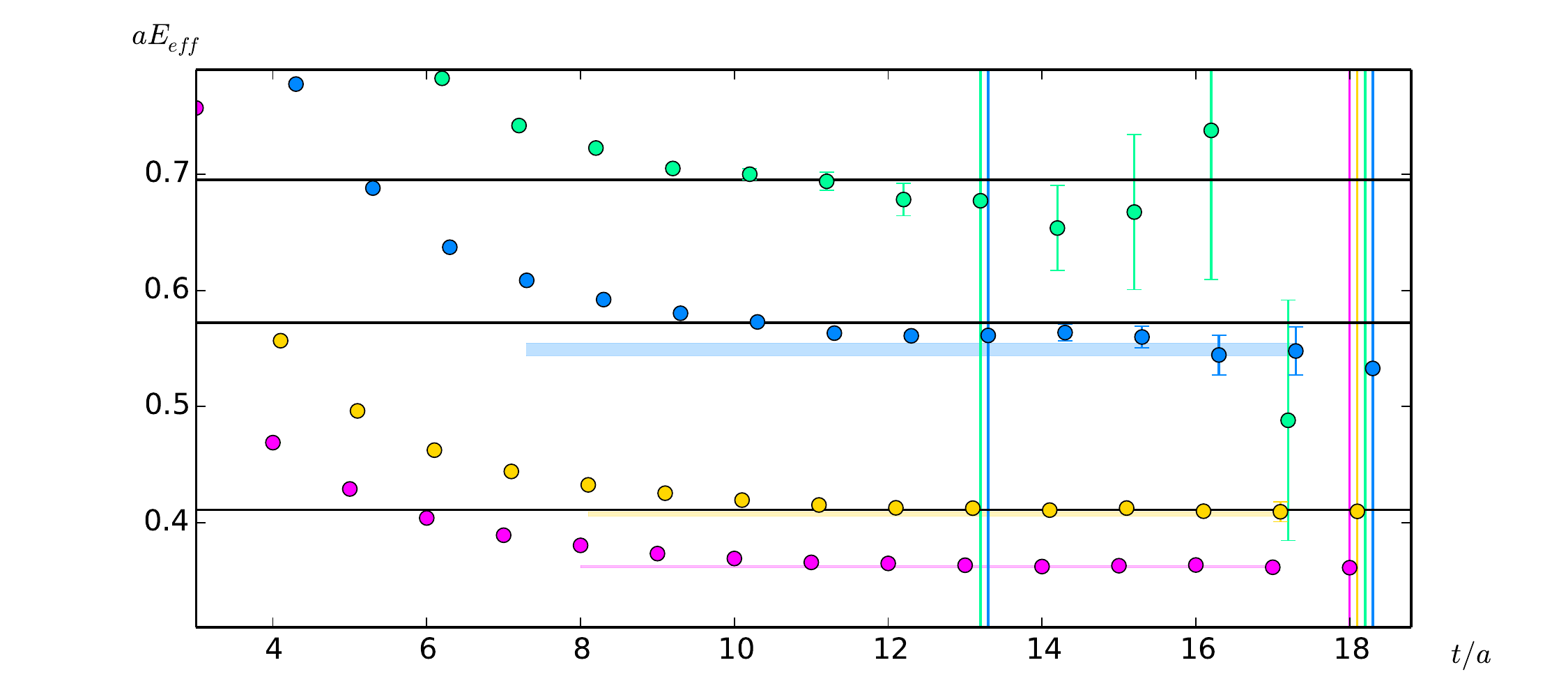}
    \includegraphics[width=.9\linewidth,clip]{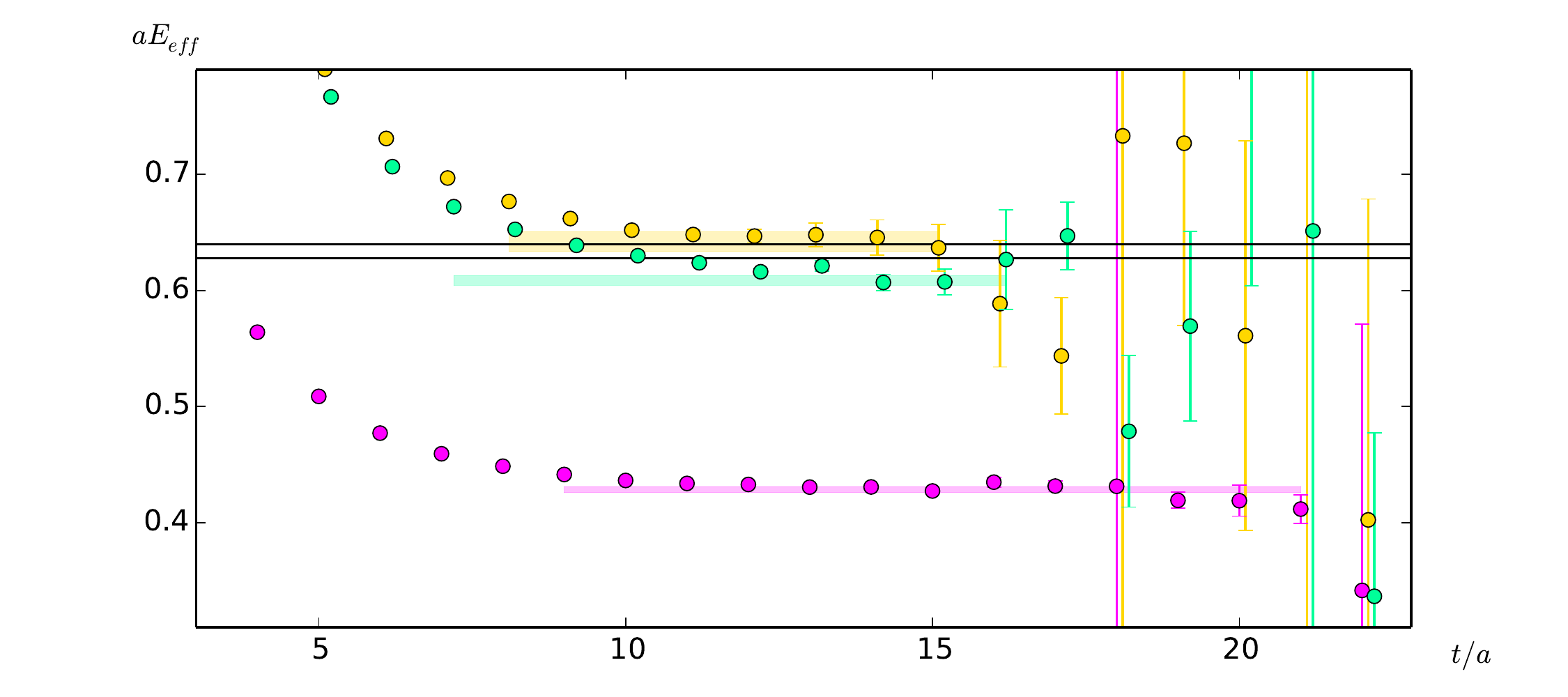}
  \caption{Effective masses of low-lying levels in the $d^2=1$, $A_1$ irrep (top) and $d^2=2$, $B_2$ irrep (bottom) on the E5 lattice.}
  \label{fig-levels-1}
\end{figure}

In each frame and irrep, we extract several energy levels in the low-lying part of the spectrum. The energy spectra of two example irreps are shown in Fig. \ref{fig-levels-1}, both of which use the window method with $t_w=3$. The solid black lines in these plots are the energies of the free pion pairs allowed in that respective box and irrep. It also shows an example of an irrep where the non-interacting levels of two moving pions are so close together that the different energies cannot be resolved with the current level of statistics.

\begin{figure}[thb] 
  \centering
  \includegraphics[width=.95\linewidth,clip]{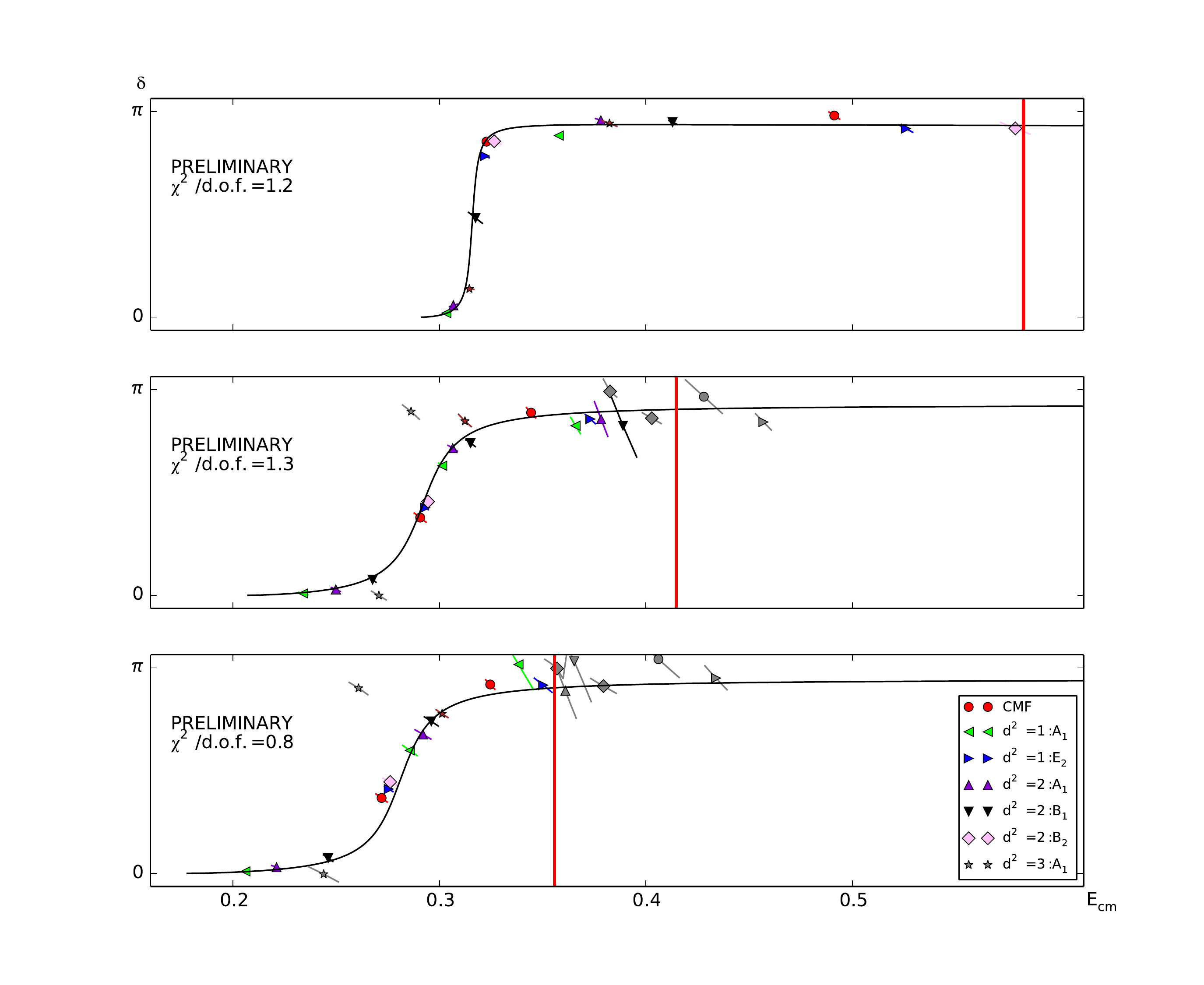}
  \caption{Energy dependence of the phase shift $\delta$ on the lattices E5 (top), F6 (center), F7 (bottom). Each data point corresponds to an energy level in an irrep relevant to the rho resonance. The resonance shape is well-described by a Breit-Wigner shape (solid line). The $4m_\pi$ threshold is indicated by the solid red line in the lower two plots. For the explanation of the grey points (not used in the fit), see text.}
  \label{fig-phase}
\end{figure}

Taking all those levels (in the range $2m_\pi \leq \sqrt{s} \leq 4m_\pi$) into account, we can map out the energy dependence of the elastic scattering amplitude parametrized by a single phase shift, plotted in Fig.~\ref{fig-phase}. The two axes in this plot are correlated, which is why the error bars of the points of this curve follow the lines allowed by the L\"uscher condition, i.e. the function $\phi(q)$. We perform a global fit to all lattice data points in this plot simultaneously using the following procedure: First of all, we assume the phase shift curve in the resonance region to follow a Breit-Wigner curve,
\begin{align}
\cot\delta_1(k) &= \frac{6 \pi}{g_{\rho\pi\pi}^2} \frac{(m_\rho^2 - E_{cm}^2) E_{cm} }{k^3} \phantom{a} .
\end{align}
The free parameters of the Breit-Wigner are the mass $m_\rho$ and the effective coupling $g_{\rho\pi\pi}$. From the L\"uscher condition we have another way of expressing $\cot\delta_1(k)$:
\begin{align*}
\delta_1(k) + \phi(q) = n\pi \Rightarrow \cot \delta_1(k) \big|_{\mathrm{L\ddot{u}scher}} \phantom{a} .
\end{align*}
The difference of these two expressions can be thought of as a function of an energy level boosted to the CMF $E_{cm}$ as well as the two resonance parameters $g_{\rho\pi\pi}$ and $m_\rho$,
\begin{align}
f(E_{cm},g_{\rho\pi\pi},m_\rho) = \cot \delta_1(k) \big|_{\mathrm{L\ddot{u}scher}} - \cot\delta_1(k,g_{\rho\pi\pi},m_\rho) \phantom{a} .
\end{align}
Given any set of parameters $g_{\rho\pi\pi}$ and $m_\rho$, the zeros of $f(E_{cm},g_{\rho\pi\pi},m_\rho)$ now correspond to energy levels $E_{cm,i}(g_{\rho\pi\pi},m_\rho)$. We can define a $\chi^2$-function using those energy levels and the ones extracted from our lattice simulation:
\begin{align}
\chi^2(g_{\rho\pi\pi},m_\rho) = \sum_{i,j} \left(E_{cm,i}(g_{\rho\pi\pi},m_\rho) - E_{\mathrm{lat},i}) C^{-1}_{i,j} 
(E_{cm,j}(g_{\rho\pi\pi},m_\rho) - E_{\mathrm{lat},j}\right) \phantom{a} ,
\end{align}
where $C^{-1}_{i,j}$ is the inverse of the covariance matrix which we have obtained from our jackknife analysis of the energy levels $E_{\mathrm{lat},i}$. By numerically minimising $\chi^2(g_{\rho\pi\pi},m_\rho)$ we can obtain the desired parameters $g_{\rho\pi\pi}$ and $m_\rho$. It is worth noting that this formalism does not depend on the fit function in the resonance region being a Breit-Wigner curve. Any other parametrisation of $\delta(k)$ could be used as well, which will be a helpful tool for the study of resonances more complicated than the $\rho$. The parameters we obtain from the fit according to our procedure on E5 are shown in Table \ref{fit-res}.

\begin{table}[thb]
  \small
  \centering
  \caption{Comparison of the Breit-Wigner fit results to the phase shift data using the two described methods.}
  \label{fit-res}
\begin{tabular}{|c|c|c|c|}
\hline 
method &  $a m_\rho$ & $g_{\rho\pi\pi}$ & $\chi^2 / d.o.f.$ \\ 
\hline 
window & 0.3161(7) & 5.94(20) & 1.2 \\ 
\hline 
fixed-$t_0$ & 0.3145(17) & 6.26(51) & 0.7 \\ 
\hline 
\end{tabular} 

\end{table}

The window method results in smaller errors with a larger $\chi^2$ value, but both methods agree with each other. For comparison, the naive $\rho$ mass on the same ensemble was determined in \cite{g-2} to be $a m_{\rho,\mathrm{naive}} = 0.3208(29)$.  

\begin{figure}[thb] 
  \centering
  \includegraphics[width=.9\linewidth,clip]{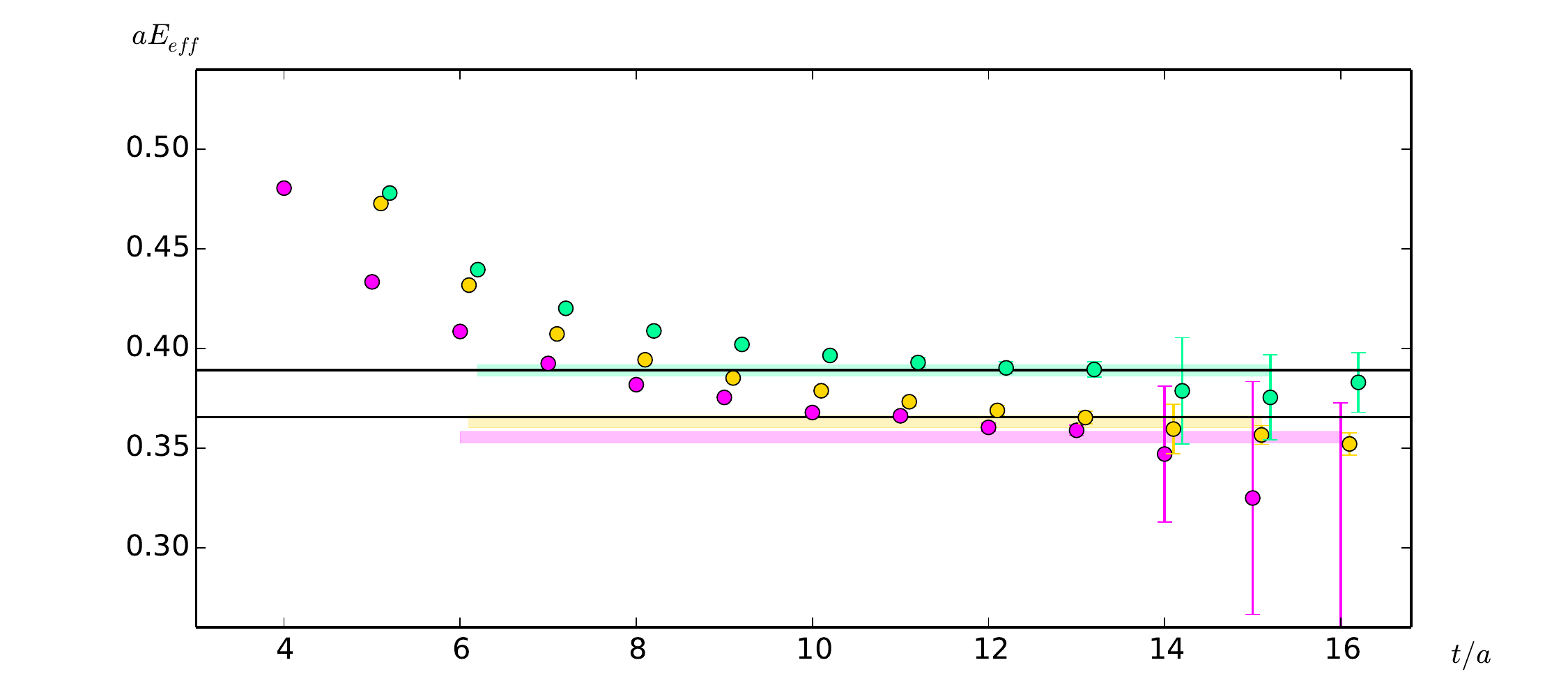}
  \caption{Effective masses of low-lying levels in the $d^2=3$, $A_1$ on the F6 lattice.}
  \label{fig-levels-3}
\end{figure}

E5 is the only lattice ensemble where we have already reached our target statistics. On F6 and F7, which are larger lattices with a pion mass of $311$ MeV and $265$ MeV, respectively, our results (see Fig. \ref{fig-phase}) look promising already, even with only a fraction of the anticipated statistics as indicated in Table~\ref{tab-ensembles}. The vertical red line indicates the $4 m_\pi$ threshold, above which the theory does not apply any more. All data above this threshold are subsequently excluded from the fit. Some other levels are excluded (indicated by a grey color of the level) because these levels are not resolvable at our current level of statistics. An example of such a situation is given in Fig. \ref{fig-levels-3}, where the lowest two levels lie so close to each other that their respective plateaus cannot be reliably determined.

\begin{figure}[thb] 
  \centering
  \includegraphics[width=.9\linewidth,clip]{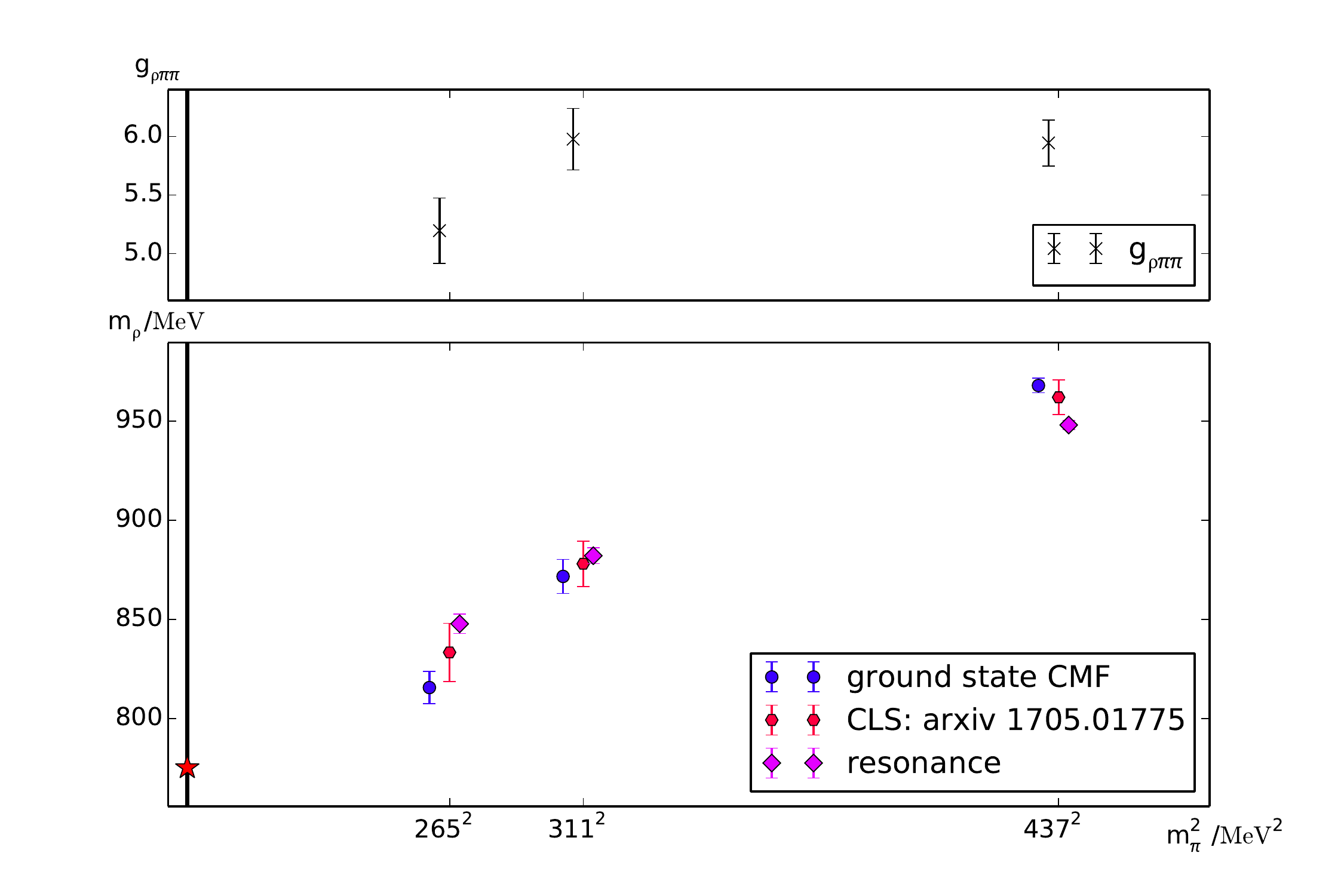}
  \caption{Pion-mass dependence of $m_\rho$ and $g_{\rho\pi\pi}$ from various determinations, using the value from the single-rho correlator at rest (blue) and from a full L\"uscher analysis taking finite-volume effects into account (magenta). For comparison, the naive rho mass determined in \cite{g-2} is shown as well (red). Grey data points are not used in the fit.}
  \label{fig-m-comp}
\end{figure}

In Fig. \ref{fig-m-comp} we show the pion-mass dependence of $m_\rho$ and the coupling $g_{\rho\pi\pi}$. It is interesting to note that an extrapolation of the naive determination of $m_\rho$, which uses the plateau value of the ground state of our correlator matrix in the centre-of-mass frame and the determination of the $\rho$ resonance mass, taking all power-law finitie-volume effects into account extrapolate quite differently towards the physical pion mass.

\section{Conclusions}\label{concl}

Disillation with stochastic LapH allows us to extract the spectrum across various irreps relevant to the $\rho$ resonance with suficcient precision to use in a L\"uscher-type analysis. While we have computed the final data set on the ensemble with the heaviest pion mass, there is still room to increase statistics further on the other two ensembles. The fit we perform to the phase shift points precisely takes into account their error, which is confined to the L\"uscher curves. Using our technology and combining it with the current matrix elements $\langle  J_\mu(t) O_i^\dagger(0) \rangle$ furthermore allows us to map out the timelike pion form factor $|F_\pi|$ in the $\rho$-resonance region, which agrees well with the Gounaris-Sakurai parametrisation of $|F_\pi|$, obtained from the phase-shift fit parameters $m_\rho$ and $g_{\rho\pi\pi}$. Finally, we use $|F_\pi|$ to constrain the large-time behaviour of the vector-vector correlator $G(t) = \langle V i (t)V i (0)\rangle$ that is used for $a_\mu^\mathrm{hvp}$. 
\newline
\newline
\textbf{Acknowledgments:}
We are grateful to our colleagues within the CLS initiative for sharing ensembles.  Our calculations were partly performed on the HPC Cluster “Clover” at the Institute for Nuclear Physics, University of Mainz.  We thank Dalibor Djukanovic for  technical  support.   We  are  grateful  for  computer  time  allocated  to project HMZ21 on the BG/Q “JUQUEEN” computer at NIC, J\"ulich.  

\clearpage
\bibliography{lattice2017}

\end{document}